\documentclass{aa}
\usepackage{graphicx}
\begin{document}
 
\title{Close frequency pairs in Delta Scuti stars}
 
\author{M.~Breger, K.~M.~Bischof}

\offprints{M.~Breger}

\institute{Institut f\"ur Astronomie der Universit\"at Wien, T\"urkenschanzstr. 17,
A--1180 Wien, Austria\\INTERNET: breger@astro.univie.ac.at}

\date{Received date; accepted date}
 
\abstract{The majority of the well-studied $\delta$~Scuti stars show 
frequency pairs in the power spectra with frequency separations less than 0.06 cd$^{-1}$ 
($0.7 \mu$Hz) as well as amplitude variability.
We examine the interpretation in terms of separate excited stellar pulsation modes,
single modes with variable amplitudes, and observational problems. The
variable-phase technique, which examines the phase jumps near the times
of minimum amplitude of an assumed single frequency, is applied to the extensive data of the star
BI~CMi, which shows some of the most extreme behavior. The following results
are found for the 5 features in the power spectrum which
could be explained as single modes with variable amplitudes
or as double modes: for three features it can be shown that
these are indeed pairs of separate
pulsation modes beating with each other: at times of minimum amplitude the phase
jumps are observed and both the observed amplitude and phase variations can be predicted
correctly by assuming two separate modes of nearly equal frequencies.
Artifacts caused by observational error,
insufficient frequency resolution or variable amplitudes can be ruled out. A fourth pair
has a probable origin in two excited modes, while a 5th case is inconclusive due to
long time scales of variability and small amplitudes.
The existence of close frequencies need to be taken into account in planning the lengths of
earth-based as well as space campaigns so that sufficient frequency resolution is obtained.
Possible reasons for the existence of close frequencies in $\delta$~Scuti stars are considered.
They include the dense frequency spacing caused by the presence of mixed modes,
rotational splitting as well as near-coincidence of the
frequencies of modes with different $\ell$ values (the so-called Small Spacing).
}
\maketitle
 
\section{Introduction}

The $\delta$ Scuti stars are variable stars situated in the Classical Instability Strip,
which pulsate with a large number of simultaneously excited radial
and nonradial modes, which makes them well-suited for asteroseismological studies. The
amplitudes of the more dominant modes in the typical $\delta$ Scuti star are a few millimag,
which is much higher than found in the Sun. It is now possible for ground-based telescopes
to detect a large number of simultaneously excited modes with millimag amplitudes
in stars other than the Sun. These studies require hundreds of telescope hours at observatories
spread around the world. 

If it can be established that $\delta$~Scuti stars contain close frequencies
of nonradial pulsation as close as 0.05 cd$^{-1}$, this has important asteroseismological
implications. Such close frequencies cannot be explained by first-order
rotational splitting since almost all $\delta$~Scuti stars rotate
faster than 10 kms$^{-1}$. The existence of extreme splitting asymmetries,
which may be produced in rapidly
rotating stars (e.~g., Goupil et al. 2000), may
produce a close doublet in the power spectrum: the observational
evidence for such theoretically predicted behavior is still missing.
Similar frequencies are also predicted in the asymptotic case
if the two modes belong to different nonradial degrees with $\ell$ values
separated by 2. In fact, the small frequency separations are
used as an important asteroseismological tool.
Furthermore, it is presently debated whether the
large number of closely spaced modes of mixed p and g character
predicted by theory are indeed found in evolved stars (e.g., see Breger \& Pamyatnykh 2002).
Close modes might provide some evidence for such modes.

It therefore appears prudent to examine whether close frequencies of pulsation
exist in $\delta$~Scuti stars and if they do, to look at their properties
in order to determine their origin.

It is important to emphasize that the presence of double peaks in the power spectrum of a pulsator,
or the emergence of a second peak after prewhitening the first peak at a similar frequency,
do not necessarily mean that there exist two pulsation modes with close frequencies. A
number of different explanations for such a behavior need to be tested. These range
from instrumental effects, methods to find the frequencies of pulsation,
to other effects intrinsic to the star. Before one can conclude the presence of
closely spaced double modes, the possible problems associated with the
available data sets need to be investigated in detail, in particular:

(i) The problem of frequency resolution: A crucial requirement
for the discovery of close frequencies is sufficient
frequency resolution. Loumos \& Deeming (1978) derive a frequency resolution
of 1.5/$\Delta$T, where $\Delta$T is the total span of observations. They make
the point that the generally used figure of 1/$\Delta$T is not really correct
from a theoretical point of view. Furthermore, data are not equidistantly
spaced and longer time gaps can occur. This can decrease
the resolution. In an extreme case, the resolution of two months
of observing spaced one year apart corresponds to that of a single month. Poor
frequency resolution, alone, does not lead to the detection of a double mode.
Nevertheless, the combination of poor frequency resolution with
one of the problems listed below, can cause severe problems in correctly interpreting
the power spectrum.

(ii) The problem of incorrect prewhitening: In the analysis
to deduce the multiple excited modes present in photometric data,
prewhitening detected frequencies is a common tool to detect further
frequencies. Prewhitening with an incorrect frequency value leaves
a spurious peak in the power spectrum with a frequency close to the prewhitened
frequency, i.~e., a doublet is found which does not correspond to two separate
modes in the star. Consequently, it is very important to
examine every detected frequency doublet whether a single frequency with a better
value can explain the doublet. 

(iii) Artifacts of amplitude variability of a single mode:
Amplitude variability on the time scale of years (or even months)
is a common feature in $\delta$~Scuti stars (e.g., see Breger 2000a,
Arentoft et al., 2001).
Some $\delta$ Scuti stars are known to exhibit very
strong amplitude variability, e.~g., the star 4~CVn with an average annual
amplitude variability of 12$\%$ for the different pulsation modes and
40$~\%$ over a decade (Breger 2000a).
The techniques used to detect the frequencies of pulsation
usually rely on constant amplitudes. The amplitude variability shows
up in the power spectrum as multiple peaks. Strictly speaking, amplitude variability
associated with a single frequency does not lead to a double peak, but to a more complex
structure. However, with amplitude variability the exact value of the pulsation
frequency can sometimes not be determined, so that the problem
of incorrect prewhitening also occurs. In some cases, with amplitude
variability only a second peak with sufficient power is noticed,
while the other features blend into the noise present in the power spectrum.

(iv) Observational problems: These range from small systematic time errors in
the data from one of the multiple observing sites to different effective wavelengths
for nominally similar filters at different sites.
The latter is caused by the fact that the amplitudes of $\delta$~Scuti
stars are extremely wavelength dependent. These problems should be carefully checked for by
the observers during data reduction, since at a later stage it is difficult to discover
such effects.

(v) Effects of aliasing: In a multiple-mode star, aliasing may result in combinations
of peaks with incorrect frequencies. Prewhitening such peaks, again, leaves artifacts
in the power spectrum, which can be interpreted as double modes. For small-amplitude
modes, we regard this effect as one of the most dangerous possibilities. This needs
to be examined carefully with statistical programs specializing in the discrimination
between various multifrequency solutions, such as PERIOD (Breger 1990) and PERIOD98
(Sperl 1998).

\section{$\delta$ Scuti stars with reported close frequencies}

A number of $\delta$ Scuti stars have been reported to have close frequency pairs with separations
less than 0.1 c/d. In order to estimate whether this is a common occurrence, we have examined the literature
for all stars listed in the $\delta$ Scuti star catalogue (Rodr\'{\i}guez et al. 2000).
For most of these stars, very little information is available, so that the question of
close frequencies cannot be addressed. Consequently, we selected those stars
which were well-studied so that close frequencies (if they exist) could be found.
Due to the inhomogeneity of the data, no rigorous, consistent definition is possible.
Therefore we selected stars with more than 200 hours of photometric observations,
annual observing runs of 20 days or longer, as well as no obvious problems of analysis, measurements,
and the distribution of the observations in time.
17 stars could be selected. Of these, seven stars
had reported close frequencies. This underestimates the true occurrence because many $\delta$ Scuti
stars have very small amplitudes so that only a few modes are observed. If we further restrict the sample
to those stars for which
more than ten frequencies of pulsation had been found, every one of the seven stars
had at least one reported
close frequency pair with separations of less than 0.06 c/d. This is surprising, since the
frequencies typically cover a wide range between 5 and 15 c/d. Due to inhomogeneity of the data
and the small number of well-studied stars, one should be careful not to overinterpret these numbers.
A safe conclusion would be that the majority of well-studied $\delta$~Scuti stars show close frequency pairs.

Table 1 lists the seven stars with reported close frequency pairs with separations of 0.06 c/d or less.

\begin{table*}
\caption[]{$\delta$ Scuti stars with reported close frequencies}
\begin{flushleft}
\begin{tabular}{lcccccl}
\hline
\noalign{\smallskip}
Star & Frequency &  Frequency & Separation & Resolution & Sufficient? &  Reference \\ 
   & (cd$^{-1}$) &  (cd$^{-1}$) & (cd$^{-1}$) & (cd$^{-1}$) & & \\ 
\noalign{\smallskip}
\hline
%\noalign{\smallskip}
\noalign{\smallskip}
 BI CMi & 4.783 & 4.818 & 0.035 & 0.003 & yes & Breger et al. (2002) \\ 
   & 8.641 & 8.658 & 0.017 & 0.003 & yes & \\
   & 10.429 & 10.437 & 0.008 & 0.003 & yes & \\
   & 12.327 & 12.350 & 0.023 & 0.003 & yes & \\ 
\noalign{\smallskip}
%\hline
%\noalign{\smallskip}
\noalign{\smallskip}
 XX Pyx & 27.011 & 27.102 & 0.091 &  0.019 & yes & Handler et al. (2000) \\
 & 38.065 & 38.110 & 0.045 & 0.019 & yes & \\
\noalign{\smallskip}
%\hline
%\noalign{\smallskip}
%\noalign{\smallskip}
 HD 18878 & 11.178 & 11.219 & 0.041 & 0.025 & yes & Mantegazza \& Poretti (1993) \\
\noalign{\smallskip}
%\hline
%\noalign{\smallskip}
%\noalign{\smallskip}
 4 CVn & 5.048 & 5.134 & 0.086 & 0.028 & yes & Breger et al. (1999)\\
  & 6.404 & 6.440 & 0.036 & 0.028 & yes  \\
  & 6.680 & 6.750 & 0.070 & 0.028 & yes & \\
\noalign{\smallskip}
%\hline
%\noalign{\smallskip}
%\noalign{\smallskip}
 FG Vir & 24.200 & 24.228 & 0.028 & 0.039 & no & Breger et al. (1998) \\ 
\noalign{\smallskip}
%\hline
%\noalign{\smallskip}
%\noalign{\smallskip}
 BV Cir & 11.077 & 11.128 & 0.051 & 0.047 & yes & Mantegazza et al. (2001) \\
 & 12.289 & 12.381 & 0.092 & 0.047 & yes & Kurtz (1981) \\ 
\noalign{\smallskip}
%\hline
%\noalign{\smallskip}
%\noalign{\smallskip}
 BW Cnc & 11.984 & 12.018 & 0.035 & 0.052 & no & Alvarez et al. (1998)\\
  & 22.516 & 22.594 & 0.078 & 0.052 & yes & Michel et al. (1999) \\
\noalign{\smallskip}						
\hline
%\noalign{\smallskip}
\end{tabular}

\end{flushleft}
\end{table*}

Let us mention two of these stars: Alvarez et al. (1998) reported
two close frequency pairs for the star BW Cnc in the
Praesepe clusters. The pairs at 138.7/139.1 and 260.6/261.5 $\mu$Hz have separations
of 0.4 and 0.9 $\mu$Hz, respectively. This translates to 0.035 and 0.078 cd$^{-1}$,
respectively. The length of the campaign of 26d suggests a frequency
resolution of 0.058 cd$^{-1}$, according to the Deeming \& Loumos criterion so that
at least one of the reported doublets has sufficient frequency resolution.
Another star is BI~CMi, which was studied extensively by the Delta Scuti Network for
a number of years so that a frequency separation as small as 0.005 cd$^{-1}$
could still be resolved.

Some of these studies are quite extensive. Due to the large amounts of data
it appears inappropriate to dismiss all reports of close frequencies
in $\delta$ Scuti stars as due to observational difficulties. It is important to
examine a star in close detail to test the different interpretations.

\begin{figure}
\centering
\includegraphics*[bb=43 472 504 745,width=86mm,clip]{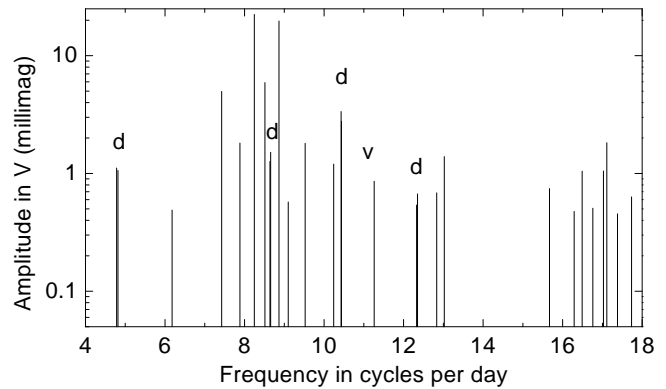}
\caption{The pulsation frequency spectrum of BI~CMi from 4 - 18 cd$^{-1}$. The double peaks in the power
spectrum are marked with a 'd', while the mode with a highly variable amplitude has been
indicated by a 'v'.}
\end{figure}

We have chosen the star BI~CMi for these detailed examinations. The choice was
motivated by the fact that the star has extensive observations available
and that we are intimately familiar with the data, since they were
obtained by the Delta Scuti Network located in Vienna.

After the pioneering study of BI~CMi by Mantegazza \& Poretti (1994), from 1997 to 2000,
the Delta Scuti Network obtained more than 1000 hours of photometric data
and deduced more than 25 frequencies of pulsation (Breger et al. 2002).
The 1998/1999 and 1999/2000 observing seasons
covered 125d and 109d, respectively. This provides a frequency resolution of 0.003 cd$^{-1}$
for the two years. One of the important and puzzling results
of the study concerns the presence
of five features in the power spectrum which can not be explained as single pulsation
modes with constant amplitudes. Their unusual nature modes was
discovered when prewhitening the frequency of each mode revealed an additional
peak in the power spectrum with a frequency very close to that of the prewhitened mode
and/or when different amplitudes were found for different years. 
We note that the detection
of these features is statistically very significant (amplitude signal/noise ratio as high as 6),
so that their reality will not be questioned here. The frequency spectrum
of BI~CMi is shown in Fig.~1, where the location of these features is also shown.

Before the nature of these features is examined in details, one should eliminate
the possibility of their origin due to observational uncertainties, complex aliasing
effects from other modes or incorrect prewhitening. While such problems can never
be eliminated completely, except for feature 5 (11.26 cd$^{-1}$) we consider them extremely
unlikely as an explanation for the following reason:
The data were  subdivided into various subgroups based on year and observatory
and a large number of new multifrequency solutions with
a variety of hypothetical scenarios computed. The result was always the same:
two modes with close frequencies or amplitude variability.
In the case of feature 5, we note the existence of a pulsation mode at 10.24 cd$^{-1}$ with a
similar amplitude, so that interaction through 1 cd$^{-1}$ aliasing presents a potential influence.
Again, no obvious alternate explanation could be derived, but for this feature
we would like to advise some caution.

Two explanations remain: pairs of close frequencies and amplitude variability. The
next section will concentrate on distinguishing between these possibilities.

\begin{table*}[t]
\caption{Parameters for the single- and double mode interpretations}
\begin{tabular}{cccrc}
\hline
\noalign{\smallskip}
   Feature & \multicolumn{2}{c}{Single-frequency model} & \multicolumn{2}{c}{Double-frequency model}\\
           &  Frequency & Amplitude in y &  Frequency & Amplitude in y \\
           &            & (millimag) &      (cd$^{-1}$) & (millimag) \\
\noalign{\smallskip}
\hline
\noalign{\smallskip}
         1 &    10.4314 & 0.6 to 6.5 &    10.4289 &        3.6 \\
           &            &            &    10.4365 &        2.7 \\
           &            &            &            &            \\
         2 &      8.6490 & 0.2 to 2.5 &     8.6578 &        1.4 \\
           &            &            &     8.6405 &        1.2 \\
           &            &            &            &            \\
         3 &     4.7826 & 0.3 to 2.7 &     4.7826 &        1.3 \\
           &            &            &     4.8179 &        1.0 \\
           &            &            &            &            \\
         4 &      12.3500 & 0.2 to 1.3 &      12.3500 &        0.8 \\
           &            &            &    12.3268 &        0.5 \\
\hline
\noalign{\smallskip}
\end{tabular}
\newline
Note that the observed amplitude range for the single-frequency model
does not correspond exactly to the sums and\\
differences of the amplitudes of the double-frequency model since the
derived parameters are based on observed data.\\   
%\end{flushleft}
\end{table*}

\section{Close frequency pairs or just amplitude variability?}

In principle, the choice between the hypotheses looks simple:
one needs to consider both models and calculate optimum fits
of the relevant parameters to the data. The models with the lowest residuals between
fit and observations would then be chosen. This valid method has
one serious drawback: in a multiperiodic star, the two models
may lead to only slightly different residuals. Furthermore, the
solution with the lowest residuals may not be the correct one due
to complex aliasing effects (see above) and the different number
of free parameters fitted.

It appears prudent to adopt a more classical, analytical approach.
The two hypotheses both lead to a variable amplitude, but to very
different phasing behavior: two beating frequencies appear as
a single frequency with variable phase. The most extreme case
occurs for two frequencies with the same amplitudes: there is 
a half cycle phase jump at every new cycle of the beat period.
Even if the amplitudes are not identical, the phase jump still exists,
but is progressively smeared out in time as the amplitudes become
more and more unequal. A numerical modeling of the amplitude ratios
and phase shifts can be seen
in Breger (1981), where the two-frequency model could be ruled out for the
cepheid HR 7308.

To distinguish between the two hypotheses, we will examine the
amplitude and phase variability for each mode of BI~CMi and compare the
one- and two-frequency models in the time, amplitude and the
time, phase domains. All computations of amplitudes and
phases of the mode to be tested for duplicity were carried out simultaneously
with all the other frequencies or frequency pairs known for BI~CMi
as listed in Breger et al. (2002). The simultaneous calculation eliminates any bias
caused by spectral leakage from other modes which were assumed to have constant annual amplitudes.
Numerical simulations showed, however, that such bias is very minor to start with.
The available data have been obtained through the Stromgren $y$ and $v$ filters.
We have used both filters to reduce the noise. For the analysis, the $v$ amplitudes were
multiplied by an experimentally determined factor of 0.632 to agree with the $y$
amplitudes and any small shifts between the times of maximum for the two filters
ignored. This is a safe procedure because in the test we are looking for
phase jumps of up to half a cycle. Furthermore, the results are insensitive
to the chosen value for the amplitude ratio because of the similar coverage
within each for year for the two colors. The data, which were obtained
at different observatories with different detectors and filters, were examined for
systematic differences (which might lead to errors in the power spectra). No
systematic effects or problems were found.

The test relies mainly on the phase shifts of the assumed single frequency.
The results can be represented in different graphical forms based on the uncertainties
of the calculated points and the number of measured data points required for
each phase range considered. Consequently, we have chosen slightly different graphical forms for
demonstrating the results for the different frequencies and to strengthen the conclusions.

Table 2 lists the four best candidates for double modes in BI~CMi and shows
the relevant parameters for both the two-frequency and single-frequency hypotheses.

\subsection{The mode(s) near 10.43 cd$^{-1}$}

The pulsation mode near 10.43 cd$^{-1}$ shows strong amplitude variability
on the timescale of about 132d. To apply the phase test we have assumed
the single (best) frequency of
10.43142 cd$^{-1}$ and computed separate solutions of the amplitudes and
phases in 10 to 15 d intervals. This could be done for all three
observing seasons. The separate results were then plotted together
utilizing a beat frequency of 132d. The results are shown in Fig. 2.

We consider the results as surprisingly unambiguous and find:

\begin{itemize}
\item
The rapid phase change is found. The fact that the phase is almost constant
outside the times of the rapid switch indicates that we have chosen a correct trial period.
The rapid phase change is incompatible with the single frequency/variable amplitude hypothesis
and a beating phenomenon of two close periods with similar amplitudes may be involved.

\item
The fast phase changes occur near the times of minimum amplitude. This is another
requirement of the two-frequency hypothesis.

\item
Both the 1998/1999 and 1999/2000 observing seasons show exactly the same
behavior. Apart from confirming our conclusions in favor of the two-frequency hypothesis
this result indicates that the amplitudes of the two pulsation modes are not
strongly variable on an annual basis. 

\item
The 7 nights of data from 1997 spanning 15 days can provide only one point in Fig. 2.
This point is in agreement with the other data and supports the two-frequency model.

\item
We have computed the best values of the two frequencies, amplitudes and phases by
a least-squares fit. In fact, the two-frequency model successfully predicts both the observed
amplitudes and phases for all three observing seasons.

\end{itemize}

\begin{figure}[!Hbt]
\centering
\includegraphics*[bb=68 56 554 731,width=86mm,clip]{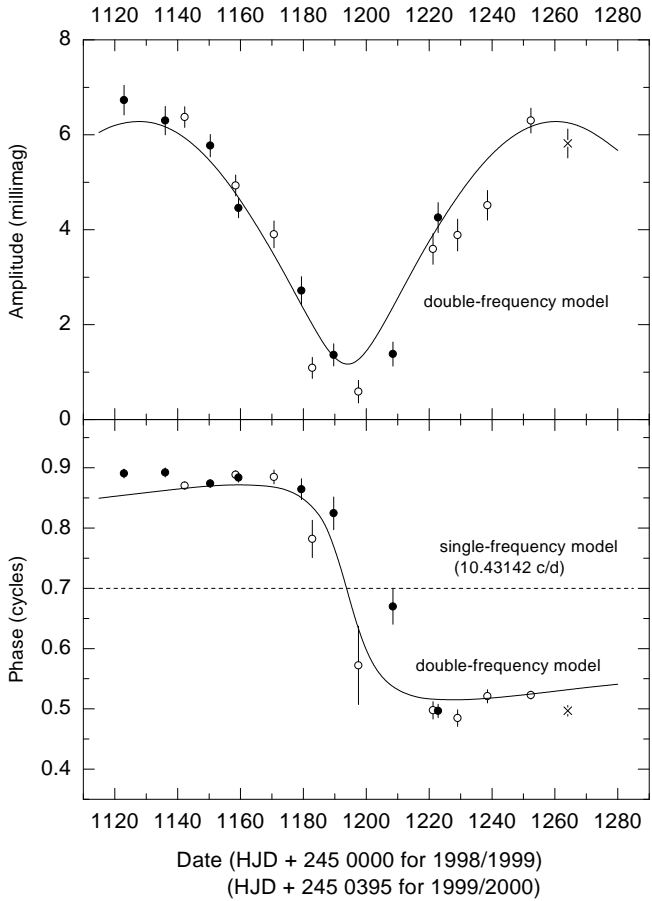}
\caption{Amplitude and phase shift diagrams to test whether the
particular mode is a single frequency with variable amplitude
or composed of two close frequencies with nearly equal amplitudes.
 Closed circles: 1998/1999 season,
open circles: 1999/2000 season, cross: 1997. The rapid phase change
in the middle of the diagram demonstrates the beating phenomenon and
the existence of two close frequencies.}
\end{figure}

We conclude that the model with a single frequency and variable amplitudes
can be ruled out from the observations and that the model with two close
frequencies provides excellent fits to the data.

\subsection{The mode(s) near 8.65 cd$^{-1}$}

The successful analysis for the previous (double) mode was repeated for the 8.65 cd$^{-1}$
mode. Here, the shorter beat cycle of $\sim$ 56d makes the test more difficult than
in the previous example, because the time bins for the examination
have to be shorter in length and thereby contain fewer points. Fig.~3 lists the
important phase information for the two extensive 1998/1999 and 1999/2000 observing
seasons. The data were plotted on the same diagram with different symbols by adopting
an experimentally determined best shift of 335d (6 beat cycles).

\begin{figure}[!Hbt]
\centering
\includegraphics*[bb=59 58 557 434, width=86mm,clip]{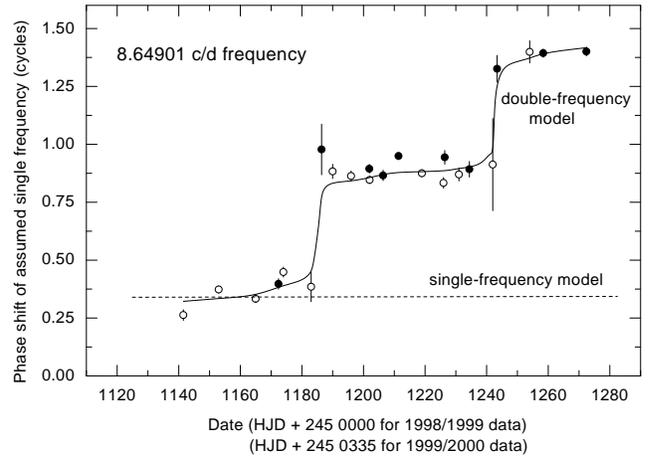}
\caption{Phase shift diagrams to test whether the
particular mode at 8.65 cd$^{-1}$  is a single frequency with variable amplitude
or composed of two close frequencies with nearly equal amplitudes. Closed circles: 1998/1999 season,
open circles: 1999/2000 season. The phase jumps demonstrate the beating phenomenon and
the existence of two close frequencies.}
\end{figure}

As for the previously examined mode, there exist regular phase jumps of
$\sim$ half a cycle. The beat cycle is short enough for each observing season
to contain two of these phase discontinuities. These jumps occur at minimum amplitude.

We conclude that the mode near 8.65 cd$^{-1}$ is actually double with two close frequencies
and that the phase jumps are incompatible with the single mode/variable
amplitude hypothesis. As was previously also found for the 10.43 cd$^{-1}$ pair, the amplitudes
associated with the two modes are constant or nearly constant on an annual basis.

\subsection{The mode(s) near 4.8 cd$^{-1}$}

The one or two modes near 4.8 cd$^{-1}$ are even more difficult to analyze because of the
short beat cycle of 28.3 d and the small amplitudes near 1 mmag. Consequently, it
was no longer feasible to examine different beat cycles separately. The 1998/1999 and 1999/2000 data
were combined. The short data set from 1997 was omitted in order to avoid potential errors
accumulated from the long time base. 

To apply the phase test we have assumed the single (best) frequency of
4.7826 cd$^{-1}$ and computed separate solutions of the amplitudes and
phases for each for the 20 phase bins.

The results are shown in Fig.~4. We can see systematic shifts in the phasing
of the assumed single frequency. These shifts correspond to the shifts calculated
from the two-frequency hypothesis. Because of the different amplitudes associated
with the two modes, a sudden phase jump is not expected in this case.

We conclude that in BI CMi there exist two close frequencies near 4.8 cd$^{-1}$.

\begin{figure}[!Hbt]
\centering
\includegraphics*[bb=78 86 558 428, width=86mm,clip]{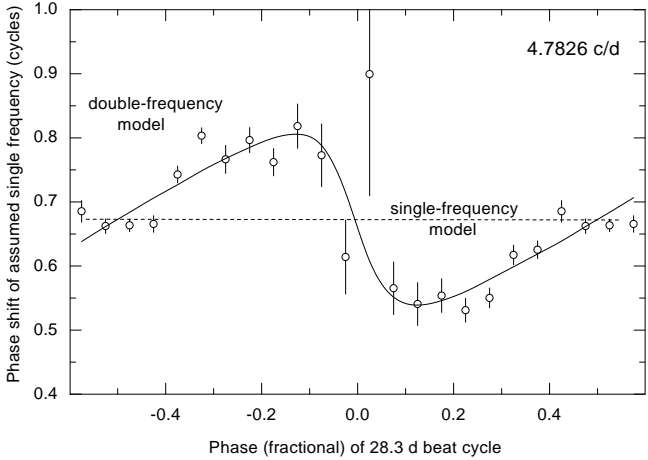}
\caption{Phase shift diagram to test whether the
particular mode at 4.8 cd$^{-1}$  is a single frequency with variable amplitude
or composed of two close frequencies with unequal amplitudes.
The phase shift of an assumed single frequency
is calculated for each phase of the beat cycle.
The large phasing uncertainty near zero beat phase
is caused by the very small amplitude at that phase.
The diagram demonstrates the beating phenomenon and
the existence of two close frequencies.}
\end{figure}

\subsection{The mode(s) near 12.35 cd$^{-1}$}

The Fourier analyses and multiple least squares solutions show two
frequencies at 12.3499 and 12.3268 cd$^{-1}$. This leads to a beat period
of 43d. As before, we have divided the data into 17 bins (each with $\sim$
1000 data points) grouped according to the phasing of the beat period.
However, because of the short beat period and the low amplitudes,
we were not able find an optimum single frequency to apply the test to.
Statistically, the best result always turned out to be one of the frequencies
of the frequency pair with a variable amplitude, as would occur by the effect
of a neighboring mode or true amplitude variability. We have chosen the
frequency with the best residuals, viz., 13.50 cd$^{-1}$. Fig.~5 shows that
the phasing test is not conclusive for this case. Both the size of the residuals
of the multiperiodic solution and the power spectra favor the two-frequency hypothesis
over the single-frequency hypothesis. Nevertheless, a convincing proof is lacking
for this frequency pair.

\begin{figure}[!Hbt]
\centering
\includegraphics*[bb=81 90 562 423, width=86mm,clip]{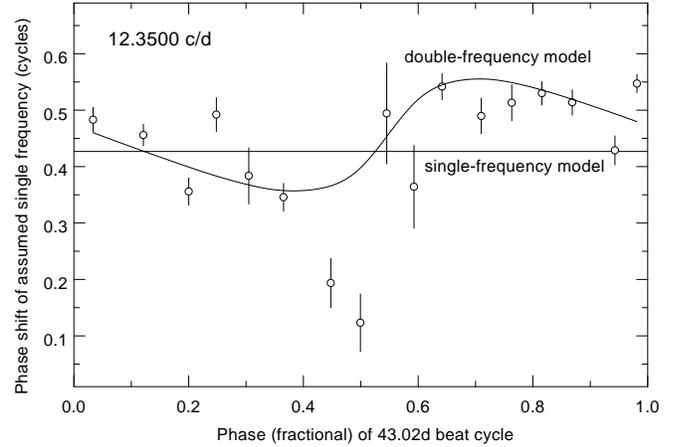}
\caption{Phase shift diagram to test whether the
particular mode at 12.35 cd$^{-1}$ is a single frequency with variable amplitude
or composed of two close frequencies. The phase shift of an assumed single frequency
is calculated for each phase of the beat cycle. The two-frequency hypothesis provides
a slightly better fit, but the scatter of the points is too large to convincingly
distinguish between the two hypotheses.}
\end{figure}

\subsection{The mode(s) near 11.26 cd$^{-1}$}

The mode at 11.2608 cd$^{-1}$ shows strong amplitude variability from year to year.
During the 1998/99 observing season, covering 120d, amplitude variability including
a minimum was found. On the other hand, no significant variability was found during
the 110d of the 1999/2000 observing season. These amplitudes, assuming a single
frequency, are shown in Table 3, where the formal uncertainties in the amplitudes
and phasing are also shown. As before, the formulae for the computation of these uncertainties assume
random errors. The true uncertainties may therefore be somewhat larger.

The amplitude variability indicates a long beat period. Extensive two-frequency
modelling shows a beat period between 500 and 600d, or $\Delta$f $\sim$ 0.002 cd$^{-1}$
between the two (hypothetical) modes. The modelling showed that the two-frequency
hypothesis is indeed a strong possibility, but that a unique, reliable
solution with two close frequencies could not be found. Including the 1991
data by Mantegazza \& Poretti ($\sim$ 2 mmag amplitude) did not improve the
situation because of the long observing gap between 1991 and 1997.

The data in Table 3 shows why the phasing test of the two-frequency hypothesis
has to be unreliable: the test relies on the phase change near minimum amplitude.
This phase change is detected, viz. 0.12 $\pm$ 0.07 cycles, but is not statistically
significant in the light of the above discussion on the uncertainties.

We conclude that due to the long beat cycle of the 11.26 cd$^{-1}$ mode,
we cannot distinguish between the variable-amplitude and two-frequency hypotheses.

\begin{table}[t]
\caption{Amplitude and phase variation of the 11.2608 mode}
\begin{tabular}{cccc}
\hline
\noalign{\smallskip}
Observing& HJD	& $y$ Amplitude	& Phasing \\
Season& 246 0000+ & (mmag) & (cycles) \\	
\hline
\noalign{\smallskip}
1997 & 2465-2481 & 2.68 $\pm$ 0.21 &	0.67 $\pm$ 0.01\\
1998/1999 & 3138-3170 &	0.64 $\pm$	0.15 & 0.61 $\pm$ 0.04\\
&3171-3190 &	0.34  $\pm$ 0.15	& 0.53 $\pm$ 0.07\\
&3191-3221 &	0.99 $\pm$  0.15 &	0.66 $\pm$	0.02 \\
&3222-3257 &	1.36 $\pm$	0.15 	& 0.64 $\pm$	0.02\\
1999/2000 & 3514-3549 &	1.68 $\pm$	0.12	& 0.62 $\pm$	0.01\\
& 3550-3583 &	1.67 $\pm$	0.12	& 0.66 $\pm$	0.01\\
& 3584-3624 &	1.67 $\pm$	0.12	& 0.65 $\pm$	0.01\\
\noalign{\smallskip}
\hline
\end{tabular}
\end{table}

\section{The strange case of 4 CVn}

We have seen that one of the signatures of two close modes
is the phase shift of up to half a cycle
near minimum amplitude. Such a phase shift has been seen before in another $\delta$~Scuti star:
the 7.37 cd$^{-1}$ mode of 4~CVn. This mode exhibits the strongest long-term amplitude variability
seen so far. Details can be found in Breger (2000a, b: Figs. 11 and 12).
The amplitude dropped from 15 mmag in 1974
to 4 mmag in 1976 and 1 mmag in 1977. In subsequent years, the amplitude has been increasing again.
The amplitude variability, by itself, is not unusual: the fascinating change is
the phase shift of 0.48 $\pm$ 0.02 cycles between 1976 and 1977, i.e., at minimum amplitude.
The determination of accurate phases became possible when the extensive 1996/7 data of 4~CVn
led to accurate frequency values, which also fit the earlier data well.

Although the half cycle shift was suggestive of a double mode, in our previous discussions of 4~CVn
we preferred the hypothesis of a single mode which was re-excited at a new epoch. The reason
was that numerical
simulations of a double mode did not lead to an improvement of the fit to the photometric data
relative to that of the single-frequency variable-amplitude hypothesis.
At this stage, we cannot present evidence proving the correctness of either the single-mode
or double-mode hypothesis, but note that in the light of the new BI~CMi results, the double-mode
hypothesis should also be kept in mind for 4~CVn. The reason for the less definite conclusions
concerning the 7.37 cd$^{-1}$ feature in 4~CVn is the relative lack of data available for the crucial 1974--1978
time period and the long time gap up to the 1983/4 data. We also note that 4~CVn contains other close
frequency pairs such as the 6.404 and 6.440 cd$^{-1}$ pair. However, if the double-mode hypothesis can be
shown to apply to the 7.37 cd$^{-1}$ feature, then the
best two-frequency solution gives a separation of $\Delta$f = 0.0015 cd$^{-1}$. This
would then be the smallest separation observed in $\delta$ Scuti stars so far.

\section{Possible reasons for close frequency pairs}

There are a number of possible explanations for the close frequency pairs:

(i) Mixed modes: Theoretical models predict a very large number of excited nonradial modes
in evolved $\delta$~Scuti stars due to the dual nature of these modes (p modes in the envelope and
g modes in the core). This dense spectrum of modes leads to small frequency spacing.
The pulsation models of 4 CVn (see Breger \& Pamyatnykh 2002) predict 554 unstable modes
with radial orders $\ell$ = 0 to 2 over a 7 cd$^{-1}$ range. Since BI~CMi is in a similar
stage of evolution, the number of modes can be regarded as an estimate for this star as well.
The average spacing of $\sim$ 0.01 cd$^{-1}$
is of the same order as the separation of the close frequencies found in this paper. If the
theoretically predicted modes are indeed unstable, the close frequencies found may find a natural
explanation.

We note here that the large number of closely spaced modes is predicted by more
than one group using independent codes, e.~g., the models for evolved $\delta$~Scuti
stars by Guzik, Bradley \& Templeton (2000) show a similar large number of predicted mixed modes.

The problem with this explanation lies in the fact that the large number of theoretically predicted
modes of mixed character have not (yet) been discovered. The $\delta$ Scuti stars select only
a small number of the possible modes to pulsate with detectable amplitudes and these
sometimes show relatively large amplitudes. It is not obvious why the (unknown) mode
selection mechanism favors so many close modes (but see below).

(ii) Trapped modes: Breger \& Pamyatnykh (2002) have explored the possibility
that the modes with low kinetic
energy are preferentially selected since they are excited more easily. These are the modes
trapped in the outer envelope and the analogs of pure acoustic modes.
In this hypothesis, only the modes with the lowest kinetic energy would be selected
out the many possible modes. The effect is very strong for $\ell$ = 1 and weaker
for $\ell$ = 2 modes. This scenario has been successful in explaining the observed $\ell$ = 1
frequency spacing in the star 4~CVn and is promising for the $\ell$ = 2 modes.
Again, the arguments could also apply to BI~CMi
because of its similar evolutionary status.

In spite of the severely reduced number of excited
low-degree modes in the trapped mode
hypothesis, two modes with close frequencies are still possible. Calculations of the
kinetic energy associated with all the possible modes show that two almost identical modes with
neighboring frequencies can have the same low kinetic energy. Both modes might, therefore,
be excited.

The average spacing of the mixed-mode hypothesis applies here as well, so that the observed frequency
separations can be matched. However, in contrast to the mixed-mode hypothesis, close frequencies
would be predicted to occur only for trapped modes, i.e., at a relatively small number of predictable
frequencies. We are, at present, engaged in testing this hypothesis with pulsation calculations.

(iii) Rotational splitting: Close frequency pairs can,
in principle, be explained by rotational splitting with only two of the
split modes visible. A simple calculation independent of the details of stellar models appears
to rule out this hypothesis. The transformation from a coordinate system rotating with the
star to the observer's coordinate system leads to a frequency separation near
$m\Omega$, where $\Omega$ is the frequency of rotation. Since almost all $\delta$~Scuti stars
rotate faster than 10 km s$^{-1}$, the predicted frequency splitting is too large.
However, this calculation may be too simple because of the extreme splitting asymmetries which can be
produced in rapidly rotating stars (e.~g., Goupil et al. 2000). Such asymmetries produce rotational splitting
which can be very much larger or smaller than that given by the $m\Omega$ term. An observational test of
this explanation and subsequent theoretical modelling requires reliable observational identification
of the the $m$ values associated with each mode. These are not available.

(iv) The Small Spacing:
Two modes with only a small frequency difference are expected if their $\ell$ values
differ by 2, at least in the asymptotic case (e.~g., Tassoul 1980). This is shown by

\begin{equation}
\nu_{n\ell} \simeq (n + \ell/2 + 1/4 + \alpha)\Delta\nu,
\end{equation}
where $n$ is the radial order and $\ell$ the degree of the pulsation mode,
$\Delta\nu$ the (large) spacing between successive orders, and $\alpha$
depends on the phase change at the stellar surface and is, in general,
a slowly varying function of frequency. This is related to the so-called Small Spacing,
which is expected to be an important asteroseismological tool for $\delta$~Scuti
stars (see Christensen-Dalsgaard 2000). The fact that many $\ell$ = 0 and 2
as well as $\ell$ = 1 and 3 pairs are separated by only very small frequency differences
could provide an explanation for the observed close frequencies.
In $\delta$~Scuti stars the pulsations are of lower radial order, so that the
asymptotic case does not really apply. Nevertheless, models show that
if this explanation for the close modes applies, the observed frequencies
would provide an important input into the pulsation models.

An important prediction of this explanation would be that the observed close
frequency pairs consist of component modes with $\ell$ values differing
by 2. Because of photometric cancellation effects across the surface, the only
pairs that we could expect to observe photometrically would be (0, 2) as well
as (1, 3). This can be checked if mode identifications are available. For the
close frequency pair in BI~CMi with the highest amplitudes, the mode identifications
from phase differences (see Breger et al 2002) suggest $\ell$ = 2 for 10.429 cd$^{-1}$
and $\ell$ = 0 (or maybe 1) for 10.437 cd$^{-1}$. The available mode identifications
for the other modes and the observed frequency spacings between the modes
(i.e., the $\Delta\nu$ value from equation 1) are at this stage not uniquely determined.
Until more data become available, the $\Delta\ell$ = 2 explanation cannot be excluded.

(v) Mode coupling:
Frequency shifts can also be produced by nonlinear mode coupling by
modes with the same or even different $\ell$ values (see Buchler, Goupil \&
Serre 1995, and Goupil 2000). These shifts could lead to close frequency pairs.
This effect has had very little observational study and additional discussion
is beyond the scope of this paper.

\section{Conclusion}

A literature search has shown that close frequency pairs with separations less
than 0.06 cd$^{-1}$ are common among well-studied $\delta$~Scuti stars. One of
the best candidates for a more detailed analysis of this phenomenon is the star
BI~CMi. The variable-phase technique, which examines the phase jumps near the times
of minimum amplitude of an assumed single frequency, was applied to test whether
theses pairs are caused by separate excited stellar pulsation modes,
single modes with variable amplitudes, or observational problems.

It was shown that at least three features are indeed pairs of separate
pulsation modes beating with each other: at times of minimum amplitude the
predicted phase jumps were observed. Both the observed amplitude and phase variations were predicted
correctly by assuming two separate modes of nearly equal frequencies.
Artifacts caused by observational error,
insufficient frequency resolution or variable amplitudes could be ruled out. A fourth pair
has a probable origin in two excited modes, while a 5th case is inconclusive due to
long time scales of variability and small amplitudes.

The existence of close frequencies needs to be taken into account in planning the lengths of
earth-based as well as space campaigns so that sufficient frequency resolution is obtained.

\acknowledgements
It is a pleasure to thank Alosza A. Pamyatnykh for many interesting discussions.
This investigation has been supported by the
Austrian Fonds zur F\"{o}rderung der wissenschaftlichen Forschung,
project number P14546-PHY.


\begin{thebibliography}{}

\bibitem{} Alvarez, M.,  Hernandez, M. M.,  Michel, E., et al. 1998, A\&A, 340, 149
\bibitem{} Arentoft, T., Sterken, C., Handler, G., et al. 2001, A\&A, 374, 1056
\bibitem{} Breger, M. 1981, ApJ, 249, 666
\bibitem{} Breger, M. 1990, Comm. in Asteroseismology (Vienna), 20, 1
\bibitem{} Breger, M. 2000a, MNRAS, 313, 129
\bibitem{} Breger, M. 2000b, in 'Delta Scuti and Related Stars',
 eds. Breger,~M., \& Montgomery, M.~H., ASP Conf. Series, Vol. 210, 3
\bibitem{} Breger, M., \& Pamyatnykh, A.~A. 2002, in 'Radial and Nonradial Pulsations as
Probes of Stellar Physics', eds. Aerts,~C., Bedding,~T., Christensen-Dalsgaard,~J.,
ASP Conf. Series, Vol. 259, 388
\bibitem{} Breger, M.,  Zima, W., Handler, G., et al. 1998, A\&A, 331, 271
\bibitem{} Breger, M., Handler, G., Garrido, R., et al. 1999, A\&A, 349, 225
\bibitem{} Breger, M., Garrido, R., Handler, G., et al. 2002, MNRAS, 329, 531
\bibitem{} Buchler, J. R., Goupil, M. J., \& Serre, T. 1995, A\&A, 296, 405
\bibitem{} Christensen-Dalsgaard, J. 2000, in 'Delta Scuti and Related Stars',
eds. Breger,~M., \& Montgomery, M.~H., ASP Conf. Series, Vol. 210, 201
\bibitem{} Goupil, M.-J., Dziembowski, W.~A., Pamyatnykh, A.~A., \& Talon,~S. 2000,
in 'Delta Scuti and Related Stars',
eds. Breger,~M., \& Montgomery, M.~H., ASP Conf. Series, Vol. 210, 267
\bibitem{} Guzik, J.~A., Bradley,~P.~A., \& Templeton,~M.~R. 2000,
in 'Delta Scuti and Related Stars',
eds. Breger,~M., \& Montgomery, M.~H., ASP Conf. Series, Vol. 210, 247
\bibitem{} Handler, G., Arentoft, T., Shobbrook, R.R., et al. 2000, MNRAS, 318, 511
\bibitem{} Kurtz, D. W. 1981, MNRAS 196, 53
\bibitem{} Loumos, G.~L., \& Deeming, T.~J. 1978, Astrophys. Space Sci., 56, 285
\bibitem{} Mantegazza, L., \& Poretti, E. 1993, A\&A 274, 811
\bibitem{} Mantegazza L., \& Poretti E. 1994, A\&A, 281, 66
\bibitem{} Mantegazza, L., Poretti, E., \& Zerbi, F.~M. 2001, A\&A 366, 547
\bibitem{} Michel, E., Hernandez, M. M., Houdek, G., et al. 1999, A\&A 342, 153
\bibitem{} Rodr\'{\i}guez, E., L\'{o}pez-Gonz\'{a}lez, M. J., \& L\'{o}pez de Coca, P. 2000,
in 'Delta Scuti and Related Stars', eds. Breger,~M., \& Montgomery, M.~H. ASP Conf. Series, Vol. 210, 499
\bibitem{} Sperl, M. 1998, Comm. in Asteroseismology (Vienna), 111, 1
\bibitem{} Tassoul, M. 1980, ApJS, 43, 469

\end{thebibliography}
\end{document}